# Flares activity of selected stars from GTSh10 catalog by CRTS data

*Gorbachev M.A.[1], Shlyapnikov A.A.[2]*

[1]*Kazan Federal University, Kazan, Russia, mark-gorbachev@rambler.ru*
[2]*FSBSI "Crimean Astrophysical Observatory RAS", Crimea, Russia, aas@craocrimea.ru*

**Abstract.** Based on the database of photometric observations of the CRTS project over an interval of more than 9 years, the activity of red dwarfs from the GTSh10 catalog was analyzed. The total number of analyzed objects, a list of stars with detected flares, amplitudes and moments of recorded events are presented. For the most interesting objects, from the point of view of the analysis of are activity, the light curves are plotted on the investigated time interval.

*Key words:* variable stars: general

## Introduction

The article presents information on the continuation of the study of are activity in red dwarfs from the GTSh10 catalog (Gershberg et al., 2011) based on the photometric observations database of the CRTS project (Drake et al. 2009). Previously, for 2032 stars from catalog GTSh10 analysis was conducted photometric observation series in band V for the period from MJD 2453464.15625 to MJD 2456591.367188 and by 360593 estimates of the brightness of 868 stars detected 2222 flares. However, further studies have shown the need for a more detailed study of the CRTS project data. Some of the discovered features of CRTS data and their correct accounting are considered in this article.

## Catalog GTSh10

In 2010, a catalog of stars with solar-type activity was prepared, which was designated GTSh10. It included dwarf stars with various manifestations of solar type activity: objects with dark spots, with hydrogen and calcium chromospheric emission, with transient flares in different wavelength ranges, with radio and X-ray radiation of star coronas (the catalog is available at: http://craocrimea.ru/~aas/CATALOGUEs/G+2010/eCat/G+2010.html).

The creation of this catalog was preceded by a lot of work done in the late 90's in the CrAO to compilation of the most complete database at that time about flaring stars and related objects.

Based on this database, which included 462 objects, it was compiled and published a paper "Catalog and Bibliography of flare stars of UV Cet - type and their related objects in the solar neighborhood" (Gershberg et al. 1999). The catalog was given in the form of an electronic supplement to the journal publication and was later printed in the monograph of R.E. Gershberg "Solar- type activity in the main sequence stars" (Gershberg 2005) and in the SIMBAD database (Wenger et al., 2006) received the designation GKL99. Currently, a new version of the Catalog of Stars with the activity of the solar type is being prepared.

## The Catalina Surveys projects and data release 2

The Catalina Surveys consist of the Catalina Sky Survey (CSS) and the Catalina Real-time Transient Survey (CRTS) NASA-funded research on observing objects near the Earth. The surveys involve three telescopes: the Mt. Lemmon Survey 1.5-m Cassegrain, Catalina Sky Survey 0.7-m Schmidt and Siding Springs Survey 0.5-m Schmidt.

CSDR2 (Catalina Surveys Data Release 2) is the second presentation of data from CRTS surveys. This data release encompasses the photometry for 500 million objects (40 billion measurements) with V magnitudes between 11.5 and 21.5 from an area of 33.000 square degrees. Catalina photometry covers objects in the range -75° < δ < 70° and |b| > 15° (more info: http://nesssi.cacr.caltech.edu/DataRelease/).

**Selection of stars for the development of processing techniques**

An analysis of the literature on the stars included in the GTSh10 catalog showed that for some of them there is, at best, one or two articles on the presence of flare activity. Based on this data, the objects were included in the catalogs of variable stars GCVS (Samus et al., 2017) and VSX (Watson et al., 2015). To confirm flare activity and refine the procedure, performed investigated of the light curves of selected objects through the photometric survey CRTS.

As a result of the analysis of 8 stars, 97 flares events were detected. Below is a method for data processing, a list of objects examined, the number of flares detected, their amplitudes and the time of registration. Table 1 contains a list of the objects investigated. The first and second columns indicate the name of the object in the GTSh10 catalog and the SIMBAD database (Wenger et al. 2000, 2006). The third and fourth columns contain the coordinates of the object at the epoch of 2000.0. In the fifth and sixth columns - the value of the average stellar magnitude according to CSDR2 in the V and Gaia DR2 (Gaia collaboration 2018) band in the G band. The seventh column contains information on the duration of observations in days (T, day) and the eighth column indicates the number of epochs of observations (Epochs). In the last column - comments.

*Table 1*

| GTSh10 | SIMBAD | R.A. (2000.0) | Decl. (2000.0) | CSDR2 | Gaia DR2 | T, day | Epochs | Note |
|---|---|---|---|---|---|---|---|---|
| 1 | 2 | 3 | 4 | 5 | 6 | 7 | 8 | 9 |
| 1062 | QR Tau | 03:47:09.01 | +22:17:32.3 | 17.98 | 18.29 | 3011 | 393 | – |
| 0972 | V1008 Tau | 03:45:30.83 | +22:33:27.8 | 15.02 | 15.26 | 3011 | 406 | – |
| 0754 | V599 Tau | 03:39:56.67 | +22:28:24.3 | 15.41 | 15.61 | 3011 | 472 | – |
| 0773 | KM Tau | 03:39:08.12 | +24:46:14.6 | 14.53 | 16.32 | 2976 | 346 | – |
| 1582 | V0697 Tau | 04:33:23.79 | +23:59:26.9 | 11.15 | 11.40 | 3121 | 360 | 1* |
| 3085 | RY Sex | 10:36:01.22 | +05:07:12.8 | 11.32 | 11.37 | 3127 | 391 | – |
| 3718 | IZ Boo | 14:20:04.68 | +39:03:01.4 | 11.28 | 11.26 | 2984 | 248 | – |
| 5367 | BD Psc | 23:02:02.98 | -02:11:39.9 | 18.05 | 18.19 | 3061 | 350 | – |

*Note: 1\* - according to the Gaia DR2 catalog, the object is a tight pair at a distance of less than 2 arc seconds with stellar magnitudes in the band G $12^m.0498$ and $12^m.2719$. The table shows the total magnitude of both components.*

Using the Gaia DR2 database (Riello et al., 2018), comparison stars were selected for each of the investigated objects by the following criteria: the bright- ness of comparison stars is comparable with the brightness of the object and the color indices of the object and comparison stars are close in value.

Also, for each object and comparison stars from the CSDR2 database, series of photometric observational data were obtained. When analysis used data for comparison stars, that coincide in time with the data for the object. Given that the filling of the CSDR2 database occurs on the basis of observations made on 3 telescopes, as well as during the implementation of this project, the recording equipment was repeatedly replaced, so the photometric material is non-uniform. As a result, there is a displacement of zero point in the calibration of stellar magnitudes. Figure 1 illustrates the light curve of one of the comparison stars in the V1012 Tau region. The graph shows that for the same object, the stellar magnitudes determined on different instruments are shown with an offset, marked with markers of different colors, this is due to the processing of observational material obtained on different instruments.

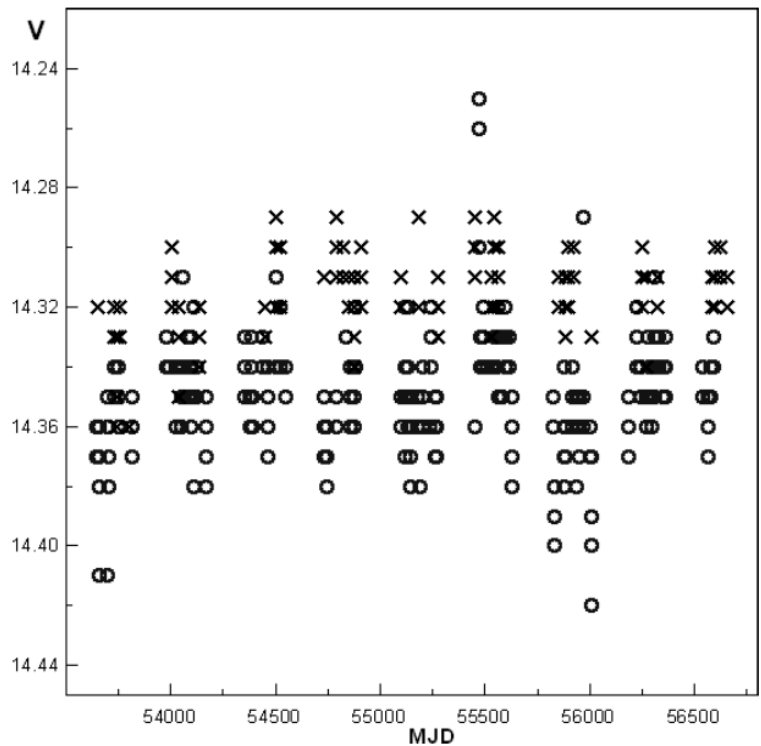

*Figure 1.* Photometric series of observations obtained on different instruments in the instrumental band V.

For the correct analysis of the flare activity of the investigated stars, the data were reduced to one zero point. This procedure was performed as follows:
- for each instrument, the mean value and the quadratic deviation were calculated;
- for the purpose of increasing accuracy when data is reduced to one zero point, from the photometric series were excluded data with large own errors and exceeding the level 3;
- star with instrumental stellar magnitude, which is close to the catalog was selected as the zero point;
- for the photometric data series obtained on other instruments, the difference between their average value and the mean value of the star selected as the zero point was calculated;
- the obtained values of the difference were added to the corresponding series of measurements.

The above described procedure for reduction the photometric series provided the possibility of a homogeneous analysis of the data obtained on different instruments.

From the consideration, time-correlated events with a gloss variation exceeding the 3 level were eliminated, since they can't be observed simultaneously at several objects. We believe that this is due to the instrumental errors of the observational series. Figure 2 shows the results of observations of the KM Tau star (open circle) and its comparison stars after the treatment described above.

To take into account the emerging trend, one of the comparison stars was selected as the reference, and all the photometric data series were reduced to a change in the brightness of this star.

Figure 3 illustrates the light curves of the KM Tau star (open circle) and the comparison star, taking into account the trend and indicating errors in the definition of stellar magnitudes according to the CSDR2 data. The axes indicate the moments of observations of stars, expressed in Julian dates and stellar magnitude.

Below in Fig. 4, the light curves of V599 Tau and comparison stars are presented, as an example of the analysis of registered low-amplitude flares.

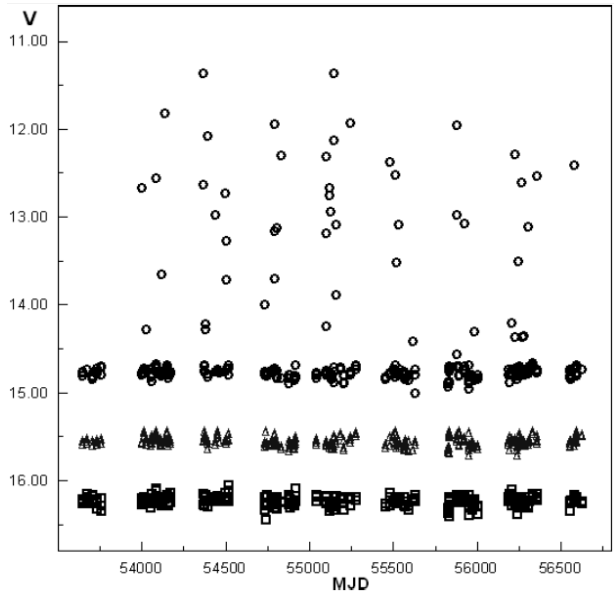
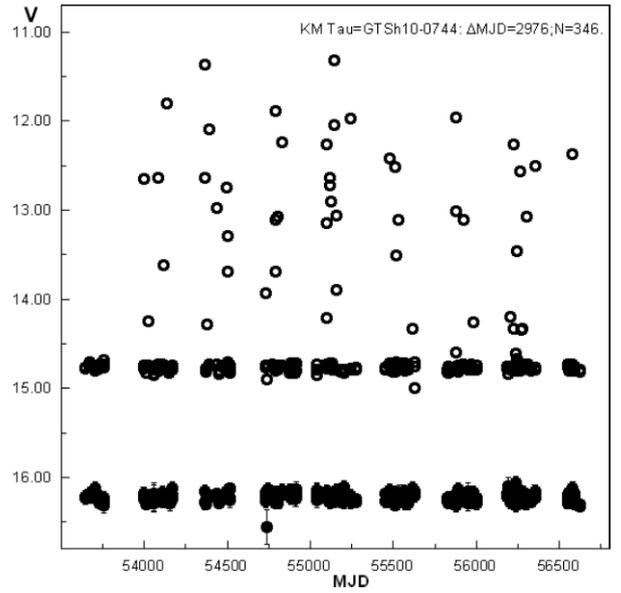

*Figure 2.* The light curves KM Tau and comparison stars after processing.

*Figure 3.* The light curves of KM Tau and the comparison star (explanations in the text).

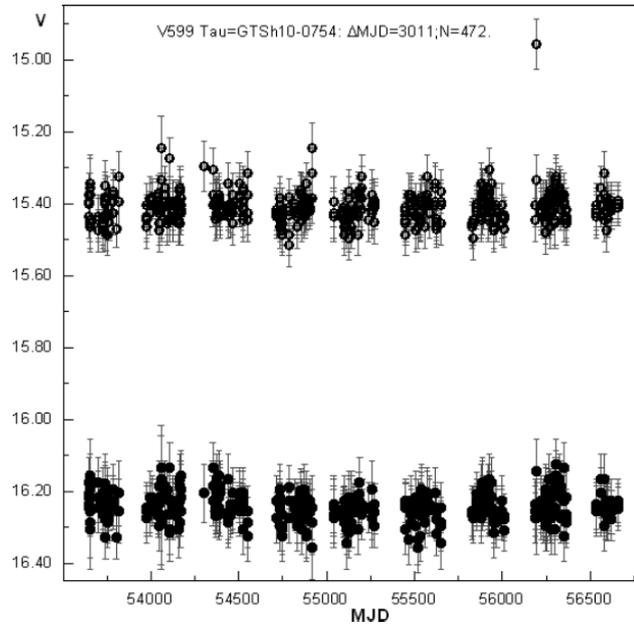

*Figure 4.* The light curves of the object V599 Tau and the comparison star.

Table 2 presents flares of 8 stars from the GTSh10 Catalog obtained from the CSDR-2 database analysis. In the first column, the flares index number is indicated, the second and third columns display the star name in the GTSh10 Catalog and the SIMBAD database. The fourth and fifth columns indicate 6 the mean value of the stellar magnitude in the V band and at the time of the flare. The sixth column shows the amplitude of the flare, expressed in stellar magnitude. The seventh column indicates the moments of flares, expressed in Julian dates.

*Table 2*

| № | GTSh | SIMBAD | $V_{mid}$ | $V_{max}$ | $\Delta V$ | MJD2400000+ | № | GTSh | SIMBAD | $V_{mid}$ | $V_{max}$ | $\Delta V$ | MJD2400000+ |
|---|---|---|---|---|---|---|---|---|---|---|---|---|---|
| *1* | *2* | *3* | *4* | *5* | *6* | *7* | *1* | *2* | *3* | *4* | *5* | *6* | *7* |
| 1 | 3085 | RY Sex | 11.344 | 10.710 | 0.634 | 53797.30798 | 5 | 3085 | RY Sex | 11.344 | 10.800 | 0.544 | 55513.53513 |
| 2 | 3085 | RY Sex | 11.344 | 10.720 | 0.624 | 53797.30822 | 6 | 3085 | RY Sex | 11.344 | 10.870 | 0.474 | 55618.43452 |
| 3 | 3085 | RY Sex | 11.344 | 10.730 | 0.614 | 53797.30869 | 7 | 3085 | RY Sex | 11.344 | 10.880 | 0.464 | 55513.53831 |
| 4 | 3085 | RY Sex | 11.344 | 10.740 | 0.604 | 53797.30845 | 8 | 3085 | RY Sex | 11.344 | 10.900 | 0.444 | 55513.53974 |

| № | GTSh | SIMBAD | $V_{mid}$ | $V_{max}$ | $\Delta V$ | MJD2400000+ | № | GTSh | SIMBAD | $V_{mid}$ | $V_{max}$ | $\Delta V$ | MJD2400000+ |
|---|------|--------|-----------|-----------|------------|-------------|---|------|--------|-----------|-----------|------------|-------------|
| 9 | 3085 | RY Sex | 11.344 | 10.970 | 0.374 | 54448.48841 | 54 | 0773 | KM Tau | 14.757 | 12.129 | 2.629 | 55143.26520 |
| 10 | 0754 | V599 Tau | 15.410 | 15.246 | 0.164 | 54057.37771 | 55 | 0773 | KM Tau | 14.757 | 11.369 | 3.389 | 55143.27242 |
| 11 | 0754 | V599 Tau | 15.410 | 15.276 | 0.134 | 54109.23623 | 56 | 0773 | KM Tau | 14.757 | 13.079 | 1.679 | 55157.27134 |
| 12 | 0754 | V599 Tau | 15.410 | 15.296 | 0.114 | 54300.46913 | 57 | 0773 | KM Tau | 14.757 | 13.879 | 0.879 | 55157.28573 |
| 13 | 0754 | V599 Tau | 15.410 | 15.306 | 0.104 | 54351.45268 | 58 | 0773 | KM Tau | 14.757 | 11.929 | 2.829 | 55241.14066 |
| 14 | 0754 | V599 Tau | 15.410 | 15.316 | 0.094 | 54549.14331 | 59 | 0773 | KM Tau | 14.757 | 12.369 | 2.389 | 55478.46743 |
| 15 | 0754 | V599 Tau | 15.410 | 15.246 | 0.164 | 54916.10929 | 60 | 0773 | KM Tau | 14.757 | 12.519 | 2.239 | 55510.31512 |
| 16 | 0754 | V599 Tau | 15.410 | 15.246 | 0.164 | 54916.10929 | 61 | 0773 | KM Tau | 14.757 | 13.519 | 1.239 | 55517.22801 |
| 17 | 0754 | V599 Tau | 15.410 | 15.316 | 0.094 | 54916.11216 | 62 | 0773 | KM Tau | 14.757 | 13.089 | 1.669 | 55532.40637 |
| 18 | 0754 | V599 Tau | 15.410 | 15.306 | 0.104 | 55927.11119 | 63 | 0773 | KM Tau | 14.757 | 14.419 | 0.339 | 55617.15120 |
| 19 | 0754 | V599 Tau | 15.410 | 14.956 | 0.454 | 56194.43174 | 64 | 0773 | KM Tau | 14.757 | 12.979 | 1.779 | 55880.37507 |
| 20 | 0754 | V599 Tau | 15.410 | 15.316 | 0.094 | 56581.34108 | 65 | 0773 | KM Tau | 14.757 | 11.959 | 2.799 | 55880.38595 |
| 21 | 0972 | V1008 Tau | 15.014 | 14.826 | 0.188 | 56302.11516 | 66 | 0773 | KM Tau | 14.757 | 13.069 | 1.689 | 55927.09747 |
| 22 | 0972 | V1008 Tau | 15.014 | 14.796 | 0.218 | 56308.13098 | 67 | 0773 | KM Tau | 14.757 | 14.309 | 0.449 | 55983.13490 |
| 23 | 1062 | QR Tau | 17.932 | 16.979 | 0.953 | 54109.22345 | 68 | 0773 | KM Tau | 14.757 | 14.199 | 0.559 | 56208.48385 |
| 24 | 1062 | QR Tau | 17.932 | 17.199 | 0.733 | 54109.23011 | 69 | 0773 | KM Tau | 14.757 | 14.369 | 0.389 | 56221.46453 |
| 25 | 1062 | QR Tau | 17.932 | 17.199 | 0.733 | 56308.16108 | 70 | 0773 | KM Tau | 14.757 | 12.279 | 2.479 | 56221.47067 |
| 26 | 1062 | QR Tau | 17.932 | 17.139 | 0.793 | 56308.17374 | 71 | 0773 | KM Tau | 14.757 | 13.499 | 1.259 | 56243.43473 |
| 27 | 5367 | BD Psc | 18.048 | 17.026 | 1.023 | 55563.10186 | 72 | 0773 | KM Tau | 14.757 | 12.609 | 2.149 | 56263.28487 |
| 28 | 0773 | KM Tau | 14.757 | 12.669 | 2.089 | 53996.44652 | 73 | 0773 | KM Tau | 14.757 | 14.369 | 0.389 | 56268.35601 |
| 29 | 0773 | KM Tau | 14.757 | 14.279 | 0.479 | 54027.36780 | 74 | 0773 | KM Tau | 14.757 | 14.349 | 0.409 | 56274.20623 |
| 30 | 0773 | KM Tau | 14.757 | 12.559 | 2.199 | 54081.35915 | 75 | 0773 | KM Tau | 14.757 | 13.109 | 1.649 | 56300.19817 |
| 31 | 0773 | KM Tau | 14.757 | 13.649 | 1.109 | 54116.26542 | 76 | 0773 | KM Tau | 14.757 | 12.529 | 2.229 | 56353.14334 |
| 32 | 0773 | KM Tau | 14.757 | 11.819 | 2.939 | 54140.16229 | 77 | 0773 | KM Tau | 14.757 | 12.409 | 2.349 | 56581.32540 |
| 33 | 0773 | KM Tau | 14.757 | 11.359 | 3.399 | 54363.45136 | 78 | 1582 | V0697 Tau | 11.063 | 10.473 | 0.590 | 55063.46941 |
| 34 | 0773 | KM Tau | 14.757 | 12.629 | 2.129 | 54363.47274 | 79 | 1582 | V0697 Tau | 11.063 | 10.313 | 0.750 | 55094.51045 |
| 35 | 0773 | KM Tau | 14.757 | 14.219 | 0.539 | 54381.34290 | 80 | 1582 | V0697 Tau | 11.063 | 10.333 | 0.730 | 55094.51097 |
| 36 | 0773 | KM Tau | 14.757 | 14.279 | 0.479 | 54381.38109 | 81 | 1582 | V0697 Tau | 11.063 | 10.333 | 0.730 | 55094.51185 |
| 37 | 0773 | KM Tau | 14.757 | 12.079 | 2.679 | 54394.37463 | 82 | 1582 | V0697 Tau | 11.063 | 10.353 | 0.710 | 55094.51264 |
| 38 | 0773 | KM Tau | 14.757 | 12.969 | 1.789 | 54438.23584 | 83 | 1582 | V0697 Tau | 11.063 | 10.363 | 0.700 | 55094.51350 |
| 39 | 0773 | KM Tau | 14.757 | 12.729 | 2.029 | 54495.16531 | 84 | 3718 | IZ Boo | 11.296 | 10.581 | 0.715 | 53900.33781 |
| 40 | 0773 | KM Tau | 14.757 | 13.269 | 1.489 | 54502.13092 | 85 | 3718 | IZ Boo | 11.296 | 10.611 | 0.685 | 53900.33892 |
| 41 | 0773 | KM Tau | 14.757 | 13.709 | 1.049 | 54502.13705 | 86 | 3718 | IZ Boo | 11.296 | 10.611 | 0.685 | 53900.34002 |
| 42 | 0773 | KM Tau | 14.757 | 13.989 | 0.769 | 54732.38765 | 87 | 3718 | IZ Boo | 11.296 | 10.571 | 0.725 | 53900.34108 |
| 43 | 0773 | KM Tau | 14.757 | 11.939 | 2.819 | 54788.21439 | 88 | 3718 | IZ Boo | 11.296 | 11.011 | 0.285 | 54593.46905 |
| 44 | 0773 | KM Tau | 14.757 | 13.699 | 1.059 | 54788.22806 | 89 | 3718 | IZ Boo | 11.296 | 10.981 | 0.315 | 55330.33782 |
| 45 | 0773 | KM Tau | 14.757 | 13.159 | 1.599 | 54788.23492 | 90 | 3718 | IZ Boo | 11.296 | 10.861 | 0.435 | 55330.34532 |
| 46 | 0773 | KM Tau | 14.757 | 13.119 | 1.639 | 54807.21603 | 91 | 3718 | IZ Boo | 11.296 | 11.001 | 0.295 | 55330.35286 |
| 47 | 0773 | KM Tau | 14.757 | 12.299 | 2.459 | 54830.20827 | 92 | 3718 | IZ Boo | 11.296 | 10.841 | 0.455 | 55636.30455 |
| 48 | 0773 | KM Tau | 14.757 | 14.239 | 0.519 | 55096.39098 | 93 | 3718 | IZ Boo | 11.296 | 10.931 | 0.365 | 55636.31100 |
| 49 | 0773 | KM Tau | 14.757 | 13.179 | 1.579 | 55102.34871 | 94 | 3718 | IZ Boo | 11.296 | 10.851 | 0.445 | 55636.31741 |
| 50 | 0773 | KM Tau | 14.757 | 12.309 | 2.449 | 55102.36709 | 95 | 3718 | IZ Boo | 11.296 | 10.911 | 0.385 | 55636.32391 |
| 51 | 0773 | KM Tau | 14.757 | 12.669 | 2.089 | 55118.33675 | 96 | 3718 | IZ Boo | 11.296 | 11.031 | 0.265 | 56371.43804 |
| 52 | 0773 | KM Tau | 14.757 | 12.759 | 1.999 | 55118.36076 | 97 | 3718 | IZ Boo | 11.296 | 11.031 | 0.265 | 56371.44197 |
| 53 | 0773 | KM Tau | 14.757 | 12.939 | 1.819 | 55126.39566 | | | | | | | |

**Conclusions**

In the course of studying the are activity of 8 stars from the GTSh10 catalog, a technique was developed and 2966 events were analyzed for more than 9 years of observations of these objects according to the Catalina Sky Survey project. 97 events are regarded by us as flares. The minimum fixed flare amplitude is $0^m.1$ of the star magnitude for V599 Tau, the maximum is $3^m.4$ for KM Tau. Given that for some of the stars investigated there was only one mention in the literature of the detection of the flares, this work confirms the presence of flare activity in these objects. When analyzing the V0697 Tau state from the data of the Gaia DR2 catalog, it was found that the object is a close pair at a distance of less than 2 arc seconds with stellar magnitudes in the band G $12^m.0498$ and $12^m.2719$.

**Acknowledge**

This work has made use of data from: the CSS survey is funded by the National Aeronautics and Space Administration under Grant No. NNG05GF22G issued through the Science Mission Directorate Near-Earth Objects Observations Program. The CRTS survey is supported by the U.S. National Science Foundation under grants AST-0909182 and AST-1313422; the SIMBAD database, operated at CDS, Strasbourg, France; the European Space Agency (ESA) mission Gaia (https://www.cosmos.esa.int/gaia), processed by the Gaia Data Processing and Analysis Consortium (DPAC, https://www.cosmos.esa.int/web/gaia/dpac/consortium). Funding for the DPAC has been provided by national institutions, in particular the institu- tions participating in the Gaia Multilateral Agreement.

The authors express their sincere gratitude to R.E. Gershberg for constant attention to the work on the flare activity of the red dwarfs and valuable comments on the text of this paper. In our research, we made use of the VizieR database and "Aladin sky atlas" developed and supported at CDS, Strasbourg Observatory, France and SAO/NASA Astrophysics Data System, USA. The authors are thankful to all who created these facilities. The first author is grateful for partial support of this work by the Russian Foundation for Basic Research. The reported study was funded by RFBR according to the research project 18-32-00775.